# Information loss and entropy production during dissipative processes in a macroscopic system kicked out of the equilibrium

**Peter Burgholzer**[1, 2]
[1] Christian Doppler Laboratory for Photoacoustic Imaging and Laser Ultrasonics, Linz, Austria
[2] Research Center for Non Destructive Testing (RECENDT), Linz, Austria
E-mail: peter.burgholzer@recendt.at

**Abstract**
In macroscopic systems behavior is usually reproducible and fluctuations, which are deviations from the typically observed mean values, are small. But almost all inverse problems in the physical and biological sciences are ill-posed and these fluctuations are highly "amplified". Using stochastic thermodynamics we describe a system in equilibrium kicked to a state far from equilibrium and the following dissipative process back to equilibrium. From the observed value at a certain time after the kick the magnitude of the kick should be estimated, which is such an ill-posed inverse problem and fluctuations get relevant. For the model system of a kicked Brownian particle the time-dependent probability distribution, the information loss about the magnitude of the kick described by the Kullback-Leibler divergence, and the entropy production derived from the observed mean values are given. The equality of information loss caused by fluctuations and mean entropy production is shown for general kicked dissipative processes from stochastic thermodynamics following the derivation of the Jarzynski and Crooks equalities. The information-theoretical interpretation of the Kullback-Leibler divergence (Chernoff-Stein Lemma) allows us to describe the influence of the fluctuations without knowing their distributions just from the mean value equations and thus to derive very applicable results, e.g., by giving thermodynamic limits of spatial resolution for imaging.

## Contents

## 1. Introduction

Many methods in non-destructive evaluation image the samples interior structure from measured signals on the surface of a sample, e.g. with ultrasound waves or by thermographic methods. The information about the interior structure has to be transferred to the sample surface, where the signals are detected and the structures are reconstructed from the measured signals. The propagation to the sample surface reduces the available information for imaging, e.g. by scattering, attenuation, or diffusion. This work was motivated by the fact that in subsurface imaging the spatial resolution is decreasing with imaging depth. *Dissipative processes* like acoustic attenuation or heat diffusion getting along with *fluctuations* of the acoustic pressure or the temperature, respectively, were identified to limit the resolution [1, 2]. In imaging there is a strong relation between spatial resolution and information content: compressed photos need less computer memory, but also the resolution gets worse. If the same structure is imaged at a higher depth, more energy of the imaging wave is dissipated before reaching the surface. This suggests that the higher *entropy production* of the imaging wave is equal to the *information loss* and a reduced spatial resolution.

Using a special Gauss-Markov stochastic process to model the diffusion of heat (Ornstein-Uhlenbeck process in Fourier k-space) this could be shown explicitly [2]. For macroscopic systems in this work it is shown for general kicked dissipative processes that the mean dissipated energy divided by the temperature is in a good approximation equal to the lost information, independent of the actual stochastic process. This is essential for application to "real-world problems" as probability distributions and stochastic processes are known only for simple physical processes or near equilibrium, where real phenomena may be approximated adequately by Gauss-Markov processes [3]. For other processes the time-dependent probability distributions are mostly unknown and thus the information loss cannot be calculated directly from the distributions, but equations for mean values are usually well known and the information loss can be estimated from the mean dissipated energy. For active thermography or photoacoustic imaging usually a short laser pulse is used to kick the sample from its equilibrium state to a state where the full spatial information is the temperature or pressure increase just after the short laser pulse [1, 2]. Due to dissipation only a part of this information can be reconstructed after a certain time from the measured temperature or pressure. Therefore the reconstructed images show less resolution and neither a better detector on the sample surface nor a subsequent signal processing algorithm can compensate this unavoidable dissipation-induced loss. This manifests a principle limit for the spatial resolution based on the second law of thermodynamics.

*Fluctuations* are deviations from the typically observed average behavior. They have been widely studied for *small systems* composed of a limited, small number of particles, as is typical for matter on meso- and nanoscales (e.g. [4,5]). *Fluctuation relations, or theorems* [6,7,8,9,10,11] describe the non-equilibrium statistical behavior of such systems. Stochastic thermodynamics [12,13,14] relates applied or extracted work, exchanged heat, and changes in internal energy along a single fluctuating trajectory. In an ensemble one gets probability distributions, and since dissipated work is typically associated with changes in entropy, one gets also a distribution for the *entropy production*.

Fluctuation relations apply on microscopic as well as macroscopic scales, but their consequences were most apparent when applied to small systems. As a system's dimensions decrease, fluctuations away from equilibrium begin to dominate its behavior. In particular, in a non-equilibrium small system, thermal fluctuations can lead to observable and significant deviations from the system's average behavior. In macroscopic systems, behavior is usually reproducible and fluctuations are small. In this work it will be shown that fluctuations play an important role also in *large systems* if chaotic behavior appears or ill-posed inverse problems have to be solved. There are numerous examples for chaotic systems in nature, where small deviations in initial conditions lead to significantly different macroscopic behavior – and even more examples exist for ill-posed inverse problems. An inverse problem is the flip side of some direct problem [15]. Inverse problems arise in practical applications whenever one needs to deduce unknown causes from observables. Direct problems treat the transformation of known causes into effects that are determined by some specific model of a natural process. Direct problems are well-posed if a unique solution exists which depends continuously on the causes, called stability [15]. As Groetsch stated almost all inverse problems in the physical and biological sciences lack these qualities and are called ill-posed [15].

The presented work helps to understand this remarkable behavior of nature as a consequence of the Second Law of Thermodynamics. Even the small fluctuations in large systems cause an *information* *loss*. We will show that for macroscopic systems this information loss is in a good approximation equal to the mean *entropy production*, defined as the mean dissipated work divided by the temperature. This information loss during the direct (forward) process cannot be compensated by any signal processing algorithm. However, additional experimental noise and insufficient data processing can discard information in addition to the unavoidable dissipation-induced loss [16]. Only if the direct (forward) process does not discard information, the inversed (reverse) process is well-posed. This is the reason for the apparent irreversibility of all but the simplest physical processes.

Mathematically the measured data is modeled as a time-dependent random variable, described by a time-dependent probability distribution. Already Claude E. Shannon connected *information* with a probability distribution [17]. The outstanding role of entropy and information in statistical mechanics was published in 1963 by E. T. Jaynes [18]. Already in 1957 he gave an information theoretical derivation of *equilibrium thermodynamics* showing that under all possible probability distributions with particular expectation values (equal to the macroscopic values like energy or pressure) the distribution which maximizes the Shannon information is realized in thermodynamics [19]. Jaynes explicitly showed for the canonical distribution, which is the thermal equilibrium distribution for a given mean value of the energy, that the Shannon or Gibbs entropy change is equal to the dissipated heat divided by the temperature, which is the entropy as defined in phenomenological thermodynamics [19, 20]. This "experimental entropy" in conventional thermodynamics is only defined for equilibrium states. By using the equality to Shannon information Jaynes recognized, that this "gives a generalized definition of entropy applicable to arbitrary nonequilibrium states, which still has the property that it can only increase in a reproducible experiment"[20].

Starting from an equilibrium state a short external pulse "kicks" the system to a *non-equilibrium state*, which evolves in time. Such a "kick" could be e.g. a short laser pulse which heats part of a sample and thus induces an acoustic [1] or a thermal wave [2]. In section 2 it is shown that for a kicked process the entropy production as the mean dissipated work divided by the temperature is equal to the information loss in a good approximation for a macroscopic system. This equality is remarkable because it is valid also *far from equilibrium* and it directly connects mean values (dissipated heat) with a statistical property of the fluctuations (information). For systems *near thermal equilibrium in the linear regime* such relations between entropy production (as the quantified dissipation) and fluctuation properties have been found by Callen [21], Welton [22] and Greene [23]. This *fluctuation-dissipation theorem* is a generalization of the famous Johnson [24] - Nyquist [25] formula in the theory of electric noise. It is based on the fact that in the linear regime the fluctuations decay according the same law as a deviation from equilibrium following an external perturbation. Section 2 is the main part of the presented work where the general conclusions for kicked processes are derived. From information theory the information loss, quantified by the Kullback-Leibler divergence, can be interpreted in the context of Chernoff-Stein's Lemma [26] to quantify the resolution limit by hypothesis testing. In section 3 these results are applied to simple model systems where the full stochastic process can be described: a kicked Brownian particle, either free (Ornstein-Uhlenbeck process) or driven by a linear or non-linear force.

The mean entropy production, defined as the mean dissipated work divided by the temperature, is equal to the information loss for any stochastic process describing macroscopic real-world phenomena with arbitrary correlations in time and space. Therefore the mean value equations also describe the influence of the fluctuations by containing the mean entropy production. The information loss, based on the Second Law of Thermodynamics, cannot be avoided. When calculated for inverse processes like imaging a principle limit of spatial resolution can be derived.

## 2. Information loss and entropy production for kicked processes

First it will be shown that for macroscopic systems the dissipated energy divided by $k_B T$ is equal to the decrease of the Kullback-Leibler divergence (KLD) $D(p_t||p_{eq})$ between the state at time t and the equilibrium state. The KLD $D(f||g)$ is used in information theory for testing the hypothesis that the two distributions with density $f$ and $g$ are different [26] and is defined as

$$D(f||g) := \int \ln\left(\frac{f(x)}{g(x)}\right) f(x) dx, \qquad (1)$$

where $ln$ is the natural logarithm; $k_B$ is the Boltzmann constant and $T$ is the temperature of the system. The second step is to show that the decrease of $k_B D(p_t||p_{eq})$ is equal to the increase of the total entropy of the system plus the surrounding heat bath, which is the Shannon information loss about the kick magnitude from the observed state at a certain time $t > 0$ after the kick. Combining these two steps for the example of a state kicked far from equilibrium one gets that by dissipating back to equilibrium the entropy production, which is the dissipated energy divided by the temperature, is equal to the information loss about the magnitude of the kick.

### 2.1. Non-equilibrium thermodynamics starting from an equilibrium state

First we follow the derivation of Kawai et al. [27] and Gomez-Marin et al. [28] (based on the Jarzynski [9] and Crooks [10,11] equalities). Jarzynski described a "forward" process starting from an equilibrium state at a temperature $T$, during which a system evolves in time as a control parameter $\lambda$ is varied from an initial value $A$ to a final value $B$. $W$ is the external work performed on the system during one realization of the process; $\Delta F = F_B - F_A$ is the free energy difference between two equilibrium states of the system, corresponding to $\lambda = A$ and $B$. The "reverse" process starts from an equilibrium state with $\lambda = B$ and evolves to $\lambda = A$ by applying the time reversed protocol for the control parameter. The superscript "tilde" refers to the corresponding time-reversed quantities. In the following, it will be useful to regard the work $W$ as a functional of the specific microscopic path trajectory followed by the system. The KLD between the forward and reverse process is [28]:

$$D(P(path)||\tilde{P}(\widetilde{path})) = \frac{\langle W \rangle - \Delta F}{k_B T} \qquad (2)$$

with the average performed work $\langle W \rangle$. The probability distribution for the forward process is $P(path)$, and $\tilde{P}(\widetilde{path})$ is the path distribution for the reverse process. The KLD multiplied by $k_B$ is equal to the average dissipated work $\langle W \rangle - \Delta F$ divided by the temperature $T$, which is the mean entropy production. Equation (2) is valid not only when the paths are in terms of microscopic variables, but also for an appropriate set of reduced variables. What is such an appropriate minimal set of variables? As stated in [28] by averaging the microscopic Crooks relation one gets $k_B D(P(W)||\tilde{P}(-W)) = (\langle W \rangle - \Delta F)/T$, which gives a surprising answer: the KLD between forward and reversed process is the same if we use all the microscopic variables or just the work $W$, which is a single scalar path-dependent variable. $D$ stays equal and does not get smaller if the set of reduced variables captures the information on the work. All the irreversibility of the process is captured by the mean dissipated work [28,37].

The Chernoff-Stein Lemma states that if $n$ data from $g$ are given, the probability of guessing incorrectly that the data come from $f$ is bounded by the type II error $\varepsilon = \exp(-nD(f||g))$, for n large [26]. In that sense $D(f||g)$ can describe some "distance" between the distribution densities $f$ and $g$. If more work is dissipated, (2) tells us that the forward and the reverse process are "more different", with $D$ as a quantitative measure of irreversibility. This result has been derived in a variety of situations such as driven systems under Hamiltonian [27,29] and Langevin [30,31,32] dynamics, as well as Markovian processes [33,34] and also electrical circuits [35,36]. Roldan and Parrondo have shown that (2) gives a useful estimate of the entropy production in a non-equilibrium stationary state (NESS) [37] and Andrieux et al. have verified it experimentally using the data of the position of a Brownian particle in a moving optical trap [31].

## 2.2. Kicked process

Instead of varying a control parameter $\lambda$ from an initial value to a final value along a given protocol as in [27] we assume to start from a canonical equilibrium state at a temperature $T$ and "kick" it at time $t = 0$. We consider a Hamiltonian $H(x)$. $x$ is a point in phase space, where $x = (q, p)$ represents the set of position and momentum coordinates. Before the kick the equilibrium probability distribution to observe the state $x$ is given by a Boltzmann distribution $p_{eq}(x) = \exp(-\beta H(x))/Z$. $Z$ is the normalization factor (partition function) and $\beta := 1/(k_B T)$. The distribution density just after the short kick is $p_{kick}(x) = p_{eq}(x - x_0)$. The performed work $W$ for a kick $x_0$ is $H(x + x_0) - H(x)$ for a phase point $x$ at $t=0$. With

$$e^{\beta W} = \frac{e^{-\beta H(x)}}{e^{-\beta H(x+x_0)}} = \frac{p_{eq}(x)}{p_{eq}(x+x_0)} \quad (3)$$

one gets by averaging with $p_{eq}$ the logarithm $ln$ of (3) and substituting $x' = x + x_0$:

$$\beta \langle W \rangle_{eq} = \int ln\left(\frac{p_{eq}(x)}{p_{eq}(x+x_0)}\right) p_{eq}(x) dx = \int ln\left(\frac{p_{eq}(x'-x_0)}{p_{eq}(x')}\right) p_{eq}(x'-x_0) dx'. \quad (4)$$

Using the definition of the KLD in (1), and that $\langle H \rangle_{kick}$ or $\langle H \rangle_{eq}$ is the average of the Hamiltonian $H$ with $p_{kick}$ or $p_{eq}$, respectively, one gets:

$$k_B D(p_{kick}||p_{eq}) = \frac{\langle W \rangle_{eq}}{T} = \frac{1}{T}(\langle H \rangle_{kick} - \langle H \rangle_{eq}) \quad (5)$$
$$\equiv mean\ entropy\ production$$

In this equation compared to (2) the distribution density $p_{kick}$ is a shifted equilibrium density corresponding to a Hamiltonian $H(x - x_0)$. The partition function $Z$ is the same for $p_{kick}$ and $p_{eq}$, therefore $\Delta F$ is zero. More performed work $\langle W \rangle$ in (5) means that the distribution $p_{kick}$ just after the kick is "more distant" from the equilibrium distribution $p_{eq}$. The equilibrium is the state where all the information about the kick magnitude is lost and all the performed work has been dissipated to the surrounding heat bath at temperature $T$.

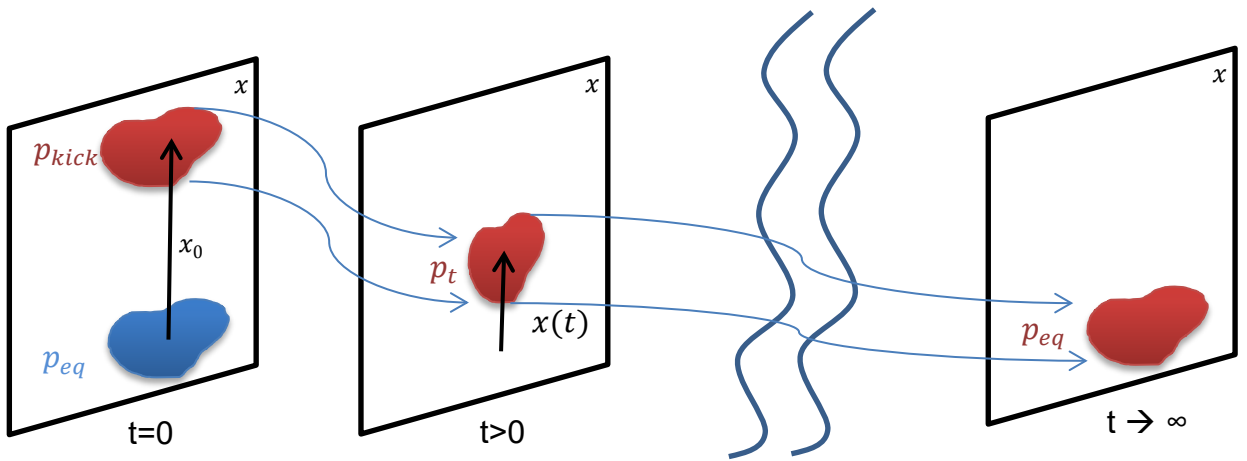

**Figure 1.** Illustration of the forward process: a system in equilibrium state $p_{eq}$ with mean value at $x = 0$ is kicked at a time $t=0$ with magnitude $x_0$ to a state $p_{kick}$ far from equilibrium, followed by a dissipative process back to equilibrium. $x$ is a set of reduced variables which captures the information on the work (see text). The arrows connecting $p_{kick}$ at time $t=0$, $p_t$ at $t > 0$, and $p_{eq}$ at $t \to \infty$ indicate the tube of trajectories, which is "thin" for macroscopic systems as deviations from the mean values $x(t)$ are small.

### 2.3. Intermediate states for the system dissipating back to equilibrium

A certain time $t > 0$ after the kick only a part of the applied work has been dissipated. To describe a state at a time $t$, which is usually not an equilibrium state, it is not necessary to know $p_t$ for all microscopic variables, but $x$ as a set of reduced variables which captures the information on the work is sufficient (see text following (2)). In Figure 1 the forward process is illustrated and the distributions are sketched. We propose that (5) can be written in a time-dependent form for all the intermediate non-equilibrium states after the kick with a distribution density $p_t$ instead of $p_{kick}$ in a good approximation.

Using the definition of the KLD in (1) and $p_{eq}$ and $p_{kick}$ from section 2.2 one gets:

$$k_B \Delta D := k_B \big(D(p_{kick}||p_{eq}) - D(p_t||p_{eq})\big)$$
$$= \Delta S + \frac{1}{T}\big(\langle H \rangle_{kick} - \langle H \rangle_{p_t}\big) = \text{mean entropy production.} \qquad (6)$$

The entropy production is the system entropy change $\Delta S$ minus the entropy flow into the heat bath, which is the negative of the dissipated heat $\langle H \rangle_{kick} - \langle H \rangle_{p_t}$ divided by the temperature. The system entropy change $\Delta S \equiv S(t) - S_{kick}$ is the difference in the Shannon entropy of $p_t$ and $p_{kick}$:

$$S(t) := -k_B \int p_t(x) \ln\big(p_t(x)\big)\,dx. \qquad (7)$$

As $p_{kick}$ is only a "shifted" equilibrium distribution $p_{eq}$, the two distributions have the same entropy: $S_{kick} = S_{eq}$.

After a long time $t$ the distribution $p_t$ converges to the equilibrium distribution $p_{eq}$ and from equation (6) one gets (5) using $\Delta S = S_{eq} - S_{kick} = 0$. This is also true in the linear regime near equilibrium as the shape of the distribution $p_t$ and therefore $S(t)$ does not change and is equal to $S_{eq}$. But also far from equilibrium for all the intermediate states $\Delta S$ is small compared to $1/T$ ($\langle H \rangle_{kick} - \langle H \rangle_{p_t}$), as for a macroscopic system fluctuations are small compared to the mean value (see Figure 1 showing a "thin" tube of trajectories). Then the distribution $p_t$ for a state far from equilibrium has nearly no "overlap" with $p_{eq}$ and one gets:

$$D(p_t||p_{eq}) = \int p_t \ln p_t\,dx - \int p_t \ln p_{eq}\,dx \approx - \int p_t \ln p_{eq}\,dx = \beta \langle H \rangle_{p_t} + \ln Z. \qquad (8)$$

The entropy term $p_t \ln p_t$ in (8) can be neglected because for all regions in the phase space where $p_t$ is different from zero and which contribute to the integral, $p_{eq}$ is nearly zero and $\ln p_t$ can be neglected compared to $\ln p_{eq}$. The same approximation in (8) is valid for $p_t = p_{kick}$ and therefore the total entropy change $\Delta S$ in (6) can be neglected:

$$k_B\big(D(p_{kick}||p_{eq}) - D(p_t||p_{eq})\big) \approx \frac{1}{T}\big(\langle H \rangle_{kick} - \langle H \rangle_{p_t}\big) \qquad (9)$$

Subtracting (9) from (5) one gets:

$$k_B D(p_t||p_{eq}) \approx \frac{1}{T}\big(\langle H \rangle_{p_t} - \langle H \rangle_{eq}\big) \approx \frac{1}{T}\big(H(x(t)) - H(x=0)\big). \qquad (10)$$

After a long time all the energy has been dissipated and $D(p_t||p_{eq})$ gets zero. The second approximation in (10) uses that for a macroscopic system fluctuations are small and the mean of the Hamiltonian is approximately the Hamiltonian of the mean value $x(t)$ (Figure 1). In section 3.3 it is shown for the example of the Brownian pendulum how good these approximations work even for non-linear equations.

The first approximation in (10) works well for macroscopic systems, as the total system entropy change $\Delta S$, which describes the influence of the change of the "shape" of $p_t$, can be neglected compared to the influence of the drift of the mean value, which is $\left(H(x(t)) - H(x=0)\right)/T$. Therefore in that approximation e.g. $p_t(x) \approx p_{eq}(x - x(t))$ can be taken having a constant "shape" all the time. Nevertheless, for the inverse problem of reconstructing the kick magnitude $x_r$ as an estimator for $x_0$ at $t=0$ from the observed value $x(t)$ at a certain time $t > 0$ after the kick with the distribution $p_t(x)$ the fluctuations around the mean value are important, which will be described in section 2.5. Before it is shown in section 2.4 that the KLD in (10) measures the information one has about the kick magnitude at time $t$.

### 2.4. Kullback-Leibler divergence as a measure of information loss

In subsurface imaging the information of the interior structure is often reconstructed by "kicking" the system out of its equilibrium. This kick can be as huge as an explosion for seismic exploration, or a short light pulse in pulse thermography (see Introduction). For reconstructions very often the wave amplitude just after the kick, named kick magnitude, is the information which should be imaged. By dissipation after some time $t$ the information about the kick magnitude is reduced and after a long time the equilibrium state is reached and all information about the kick magnitude is lost (see Figure 1). The Shannon entropy $S(t)$ (7) of the kicked system is no suitable measure for this information, as it stays approximately constant. One has to take into account in addition the entropy of the surrounding heat bath. The heat bath is a big heat reservoir compared to the dissipated heat. Therefore its temperature $T$ does not change and its entropy increase $\Delta S_r$ (reservoir) is the dissipated heat divided by $T$. The information loss is the increase in total entropy $\Delta S_{tot}$, which turns out to be equal to the difference of the Kullback-Leibler divergence (KLD) from (6):

$$\text{information loss} = \Delta S_{tot} = \Delta S + \Delta S_r = \Delta S + \frac{1}{T}\left(\langle H \rangle_{kick} - \langle H \rangle_{p_t}\right) = k_B \Delta D \qquad (11)$$

For the transient relaxation to the equilibrium state it was already shown by van Kampen [38] and later e.g. by Esposito et al. ([39], [40]) that the KLD, also called the relative entropy, measures the change in total entropy. In information theory, the KLD $D(p_t || p_{eq})$ in (10) can be identified as the amount of information that needs to be processed to switch from the known equilibrium distribution $p_{eq}$ (no kick) to the distribution $p_t$ a time $t$ after the kick [26]. If in the definition (1) of the KLD the logarithm to the base 2 is taken instead of the natural logarithm, the KLD measures the average number of bits needed to describe the kick magnitude, if a coding scheme is used based on the given distribution $p_{eq}$ rather than the "true" distribution $p_t$ [26].

There exists a generalization of the second law and of Landauers principle for states arbitrarily far from equilibrium given by Hasegawa et al. ([41], [42]) and later by Esposito and Van den Broeck in [43]. The main idea to deal with a non-equilibrium state $p_t$ is to perform a sudden quench from the known Hamiltonian $H$ to a new one $H^*$, such that the original non-equilibrium state becomes canonical equilibrium with respect to $H^*$. The average amount of irreversible work for this quench turns out to be the KLD $D(p_t || p_{eq})$ times $k_B T$. In that sense $k_B D(p_t || p_{eq})$ is the information needed to specify the non-equilibrium state $p_t$. Alternatively, one can say that the measurement at time $t$ has decreased the total entropy by $D(p_t || p_{eq})$. The processing of the gained information from the measurement by a physical device will at least offset the decrease of total entropy that was realized in the measurement.

Equation (10) is the first main result and states that for a macroscopic system the information $k_B D(p_t || p_{eq})$ about the non-equilibrium state any time $t$ after the kick is just the mean work, which has not been dissipated yet, divided by the temperature $T$.

## 2.5. Inverse problem, time reversal and spatial resolution

The magnitude of the kick $x_0$ should be estimated from the observed value $x$ at a certain time $t > 0$ after the kick. To find the distribution $p_r$ of the estimated $x_r$ ($r$ named for *r*econstructed) a function $f$ is defined, which describes the temporal change of the mean value $x(t)$ from time zero to time $t$: $x(t) \equiv f(x_0)$. The time reversal of the mean value equation is the inverse function $f^{-1}$ with $x_0 = f^{-1}(x(t))$. Then the distribution density for the estimated kick magnitude is:

$$p_r(x_r) = p_t(f(x_r)) \left| det\left(\frac{\partial f}{\partial x}(x_r)\right) \right|, \tag{12}$$

with $det$ is the determinant of the Jacobian matrix $\partial f / \partial x$. The standard deviation for $x_r$, which is the square root of the variance of $p_r$, can be taken as a measure of the spatial resolution $\delta_i$ in the direction of the $i$-th component of $x$. For dissipative processes, as sketched in Figure 1, after a long time equilibrium is reached. For the inverse problem small fluctuations described by $p_t$ are highly amplified and the inverse problem gets ill-posed. The function $f$ is not bijective anymore and regularization methods, which make additional assumptions, have to be used to find a good estimate for the magnitude of the kick $x_r$.

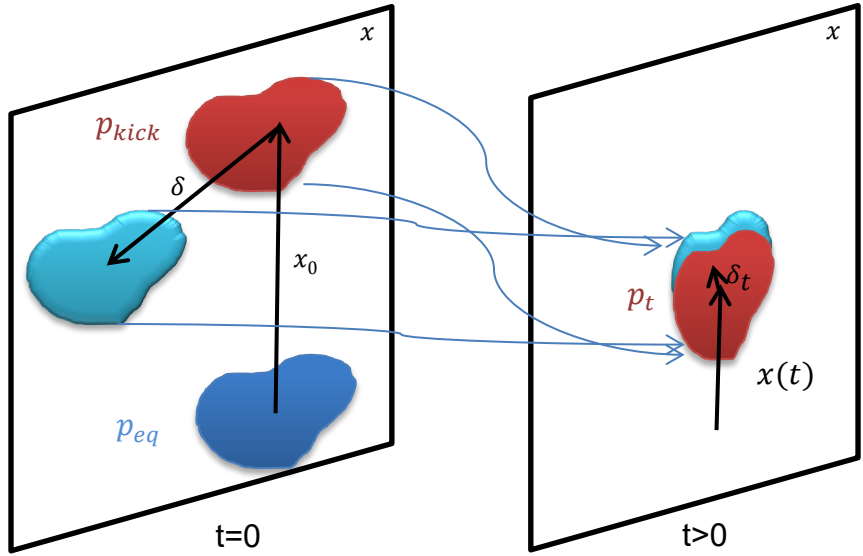

**Figure 2.** Illustration of resolution: the distribution density $p_{kick}$ just after the kick ($t = 0$) is the distribution $p_{eq}(x - x_0)$ and $p_{eq}(x - x_0 - \delta)$, respectively. With ongoing time this two distributions overlap more and more, which can be quantitatively described by a decreasing KLD $D(p_t(\delta)||p_t(\delta = 0))$. The resolution is the smallest $\delta$ for which these two distributions can be still distinguished at a time t. The type II error $\varepsilon$ to take the wrong distribution is given by the Chernoff-Stein Lemma.

In this section the resolution of the estimated kick magnitude as a function of time $t$ after the kick is derived by using the information theoretical meaning of $D$ (Chernofff-Stein Lemma) as a "distance" between two distributions. The resolution $\delta \equiv (0, ..., 0, \delta_i, 0, ..., 0)$ is the smallest difference in the kick magnitude $x_0$ and $x_0 + \delta$ which can be detected after the time $t$ in the component i of the set of reduced variables. The distribution density just after the kick is the distribution $p_{eq}(x - x_0)$ and $p_{eq}(x - x_0 - \delta)$, respectively (Figure 2, left for $t = 0$). After some time $t > 0$ the "distance" $\delta_t$ of the two distributions gets less, which can be quantitatively described by the decreasing KLD $D(p_t(\delta)||p_t(\delta = 0))$ (Figure 2, right for $t > 0$). Similar to the derivation of (10) the decrease in $D$ can be calculated from the mean dissipated work. For the time $t = 0$ following the ideas for the derivation of (5) in (3) and (4) one gets from $p_{eq}(x) = \exp(-\beta H(x))/Z$:

$$k_B D(p_{eq}(x - x_0 - \delta)||p_{eq}(x - x_0)) = k_B \int \ln\left(\frac{e^{-\beta H(x-x_0-\delta)}}{e^{-\beta H(x-x_0)}}\right) p_{eq}(x - x_0 - \delta)\, dx =$$
$$= \frac{1}{T}\int (H(x' + \delta) - H(x')) p_{eq}(x')dx' = \frac{1}{T}(\langle H\rangle_\delta - \langle H\rangle_{eq}) \equiv \frac{\langle W_\delta\rangle}{T} \quad (13)$$
$$\approx \frac{1}{T}(H(\delta) - H(\delta = 0)).$$

To get from the first line of (13) to the second line we have substituted $x' = x - x_0 - \delta$. Using the same arguments for neglecting the influence of the "shape" change of $p_t$ in $\Delta S$ as in (10), now (13) can be written in a time-dependent form for the states $p_t$ after the kick:

$$k_B D(p_t(\delta)||p_t(\delta = 0)) \approx \frac{1}{T}(\langle H\rangle_{\delta_t} - \langle H\rangle_{eq}) \equiv \frac{\langle W_\delta(t)\rangle}{T}$$
$$\approx \frac{1}{T}(H(\delta_t) - H(x = 0)). \quad (14)$$

According the Chernoff-Stein Lemma (last paragraph in section 2.1) the error $\varepsilon$ to take the wrong distribution is $\exp(-nD)$, for n a large number of data points, which gives:

$$\langle W_\delta(t)\rangle \approx k_B T \frac{1}{n} \ln\left(\frac{1}{\varepsilon}\right). \quad (15)$$

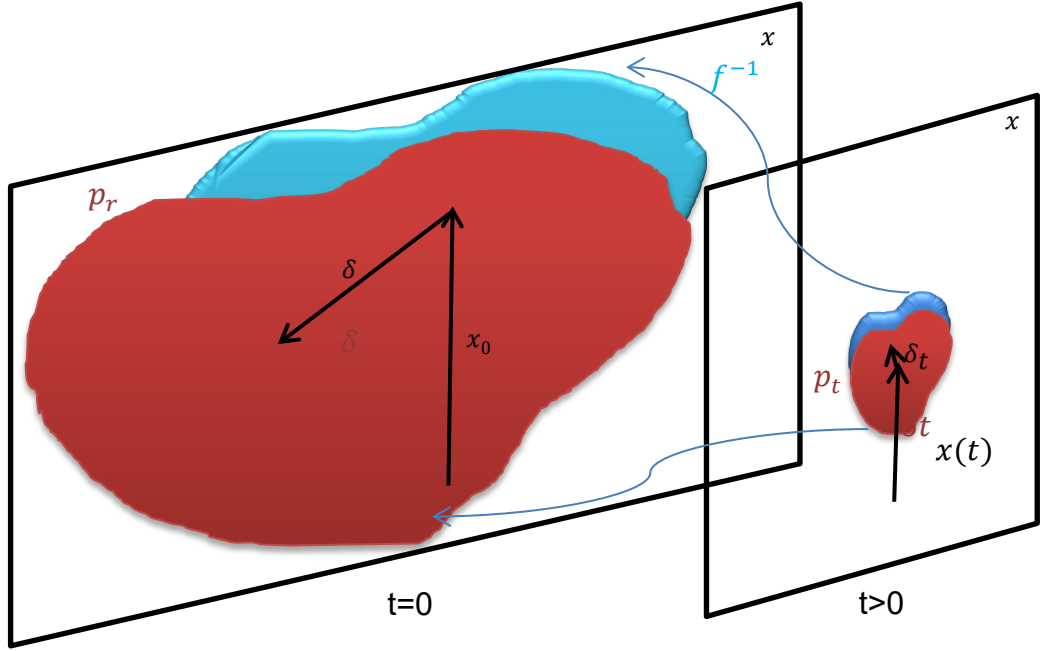

**Figure 3.** Illustration of the inverse problem: from the observed value $x(t)$ at a certain time $t > 0$ after the kick with the distribution $p_t$ the magnitude of the kick should be estimated, which is an ill-posed problem and the fluctuations are highly "amplified". The distance of the $\delta$-shifted distribution from the un-shifted distribution at time $t > 0$ (right) is the same as for the reconstructed kick-magnitudes $x_r$ with distribution density $p_r$ at $t=0$ (left) (16). The distance is quantified by the KLD.

In Figure 3 the inverse problem of Figure 2 is sketched. The decreasing distance $D(p_t(\delta)||p_t(\delta = 0))$ for the forward process results in a big amplification of the fluctuations for the reversed process, shown in the reconstructed probability density $p_r$ at $t=0$. The distance between the $\delta$-shifted distribution and the un-shifted distribution is the same for the reconstructed densities ($t=0$, Figure 3 left) and at $t > 0$ (Figure 3, right), which can be seen by using (12) and substituting in the integral (1) in the definition of the KLD $x = f(x_r)$ and $dx = \left|\det\left(\frac{\partial f}{\partial x_r}\right)\right| dx_r$:

$$D(p_r(\delta)||p_r(\delta = 0)) = \int \ln\left(\frac{p_r(\delta; x_r)}{p_r(\delta = 0; x_r)}\right) p_r(\delta; x_r)\, dx_r$$
$$= \int \ln\left(\frac{p_t(\delta; x)}{p_t(\delta = 0; x)}\right) p_t(\delta; x)\, dx = D(p_t(\delta)||p_t(\delta = 0)). \tag{16}$$

Therefore fluctuations play an important role also for macroscopic systems if ill-posed inverse problems are involved. But the time evolution of these fluctuations is determined by the mean value equations if the variables capture the information on the dissipated work and no elaborated methods for stochastic processes have to be used. Such methods will be used in the next section for the study case of a kicked Brownian particle and compared with results from simple mean value equations.

## 3. Study cases for a kicked Brownian particle

We now consider the velocity $v$ of a particle as a stochastic process, but for simplicity only one velocity component (one dimension). Stochastic processes can be described mathematically e.g. by Master equations, Langevin- or Fokker-Planck equations, like in the books of van Kampen [38], Gardiner [44], or Risken [45]. In the Langevin-equation the environmental forces on a particle in Newton's law are a linear damping term together with random noise:

$$\frac{dv(t)}{dt} = -\gamma v(t) + \sigma \eta(t). \tag{17}$$

The linear damping $\gamma v$ is a viscous drag and $\sigma$ is the amplitude of the white noise $\eta$, which has a zero mean value and is uncorrelated in time: $\langle \eta(t)\eta(t') \rangle = \delta(t-t')$ with the Dirac delta function $\delta(t)$. The Langevin equation governs an Ornstein-Uhlenbeck (O-U) process, named after L. S. Ornstein and G. E. Uhlenbeck, who formalized the properties of this continuous Markov process [46]. It was shown by using the statistical properties and the continuum limit of the white noise $\eta$, that the Langevin equation is equivalent to a description based on a Fokker-Planck equation for the time-dependent distribution density $p(v,t)$ of the velocity [e.g. 4, 45]:

$$\frac{\partial p(v,t)}{\partial t} = \frac{\partial(\gamma v p(v,t))}{\partial v} + \frac{\sigma^2}{2}\frac{\partial^2 p(v,t)}{\partial v^2}. \tag{18}$$

Usually a certain initial velocity $v_0$ at time $t=0$ is chosen [e.g. in 4], which gives for $p(v,t)$ a Gaussian distribution with time-dependent mean and variance. We start with the equilibrium state (zero time derivative if inserted into (18)) as the initial velocity distribution according to Jarzynski as described in section 2.1:

$$p_{eq}(v) = \frac{1}{Z}\exp(-\frac{\gamma}{\sigma^2}v^2). \tag{19}$$

At time zero the particle is kicked, which causes an immediate change in velocity of $v_0$ (kick magnitude) and the distribution density after the kick is

$$p_{kick}(v) = p_{eq}(v-v_0) = \frac{1}{Z}\exp(-\frac{\gamma}{\sigma^2}(v-v_0)^2). \tag{20}$$

The solution of (18) for the time-dependent distribution density $p_t(v)$ at $t > 0$ with $p_{kick}(v)$ as initial condition at $t = 0$ is

$$p_t(v) = \frac{1}{Z}\exp(-\frac{\gamma}{\sigma^2}(v-\bar{v}(t))^2) \; with \; \bar{v}(t) = v_0 \exp(-\gamma t) \tag{21}$$

which gives a Gaussian distribution with time dependent mean value $\bar{v}(t)$ but a constant variance $Var(v) = \sigma^2/(2\gamma)$. As shown in [2] this is a general feature of Gauss-Markov processes, also for higher dimensions: taking the equilibrium as an initial condition results for all times after the kick in a distribution with a constant (co)variance (matrix) equal to the equilibrium variance. One realization of the kicked O-U process is shown in Figure 4. At the time $t = 0$ a kick with a magnitude of $v_0 = 10$ occurs. For a time $t > 0$ the information about the magnitude of the kick gets more and more lost due to the fluctuations. In the next subsection this increasing information loss is quantified and compared to the mean entropy production.

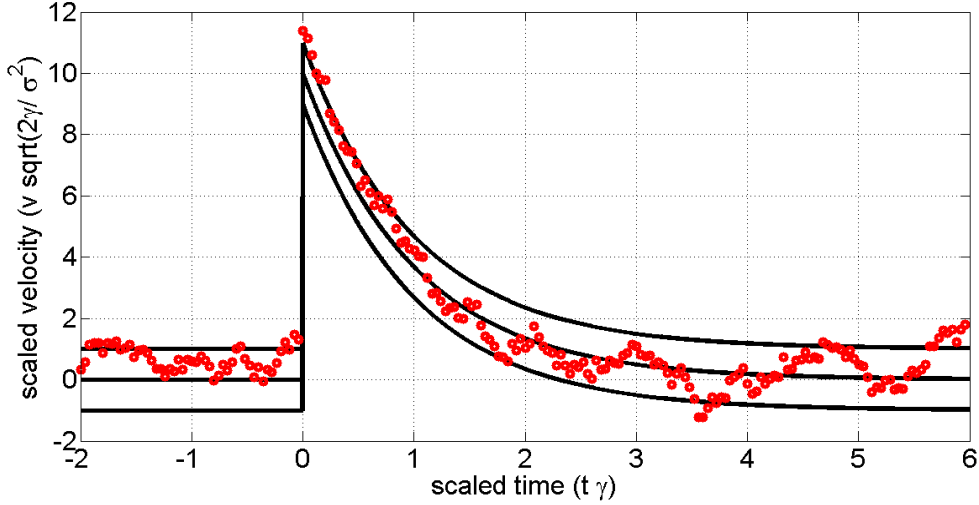

**Figure 4.** The red circles show a typical realization of a kicked Ornstein-Uhlenbeck process defined by the Langevin equation (17). The scaled time $t\gamma$ is on the horizontal axis. The velocity $v$ on the vertical axis is scaled to have a unit variance. At the time $t = 0$ a kick magnitude of $v_0 = 10$ has been added to the scaled velocity. Increasing in time the information about the magnitude of the kick gets more and more lost due to the fluctuations. The solid lines represent the mean, and mean ± standard deviation, which is the square root of the variance.

### 3.1. Ornstein-Uhlenbeck process

The main question of this subsection is: how well can we estimate the magnitude $v_0$ of the kick at $t = 0$ from a measurement of the velocity $v(t)$ at a time $t > 0$ ? The information content of the time-dependent velocity distribution density $p_t(v)$ given in (21) is

$$k_B D(p_t(v) || p_{eq}(v)) = k_B \frac{\gamma v_0^2}{\sigma^2} e^{-2\gamma t}, \quad (22)$$

which shows an exponential decay and therefore a loss of information. After a long time $t$, in the equilibrium state, all the information is lost and the mean velocity $\bar{v}(t)$ is zero again, as it was before the kick. Then all the applied work for the kick, which is the kinetic energy $H(v) = \frac{1}{2}mv^2$ of the Brownian particle with mass $m$, has been dissipated. For the Brownian particle equation (5) reads as:

$$information\ loss = k_B D(p_{kick} || p_{eq}) = k_B \frac{\gamma v_0^2}{\sigma^2} =$$
$$= mean\ entropy\ production = \frac{\langle W \rangle_{eq}}{T} = \frac{mv_0^2}{2T}, \quad (23)$$

which gives for the variance of the velocity distribution:

$$Var(v) = \frac{\sigma^2}{2\gamma} = \frac{k_B T}{m}. \quad (24)$$

This relation states a connection between the strength of the fluctuations, given by $\sigma^2$, and the strength of the dissipation $\gamma$. This is the fluctuation-dissipation theorem (section 1) in its simplest form for uncorrelated white noise and this derivation of (24) shows its information theoretical background. In the past (24) has been derived for equilibrium by using the equipartition theorem, which states that the equilibrium energy associated with fluctuations in each degree of freedom is $k_B T/2$.

The time-dependent equation (10) reads for the kicked O-U process for times after the kick as (using (22) and (24)):

$$k_B D(p_t(v) || p_{eq}(v)) = \frac{1}{T} \frac{mv_0^2}{2} e^{-2\gamma t} = \frac{1}{T}\big(H(\bar{v}(t)) - H(\bar{v} = 0)\big). \quad (25)$$

The approximations in (10) are fulfilled exactly. Because of the linearity of the Langevin-equation (17) the "shape" of $p_t$ does not change and therefore the system entropy change $\Delta S$ stays exactly zero for all times $t$.

The mean kinetic energy of a Brownian particle decreases by a factor $\exp(-2\gamma t)$, giving for the velocity resolution $\delta_v$ using (15):

$$\langle W_\delta(t) \rangle = \frac{m\delta_v^2}{2} e^{-2\gamma t} \approx k_B T \frac{1}{n} \ln\left(\frac{1}{\varepsilon}\right) \Rightarrow \delta_v^2 \approx \frac{2k_B T}{mn} \ln\left(\frac{1}{\varepsilon}\right) e^{2\gamma t}. \qquad (26)$$

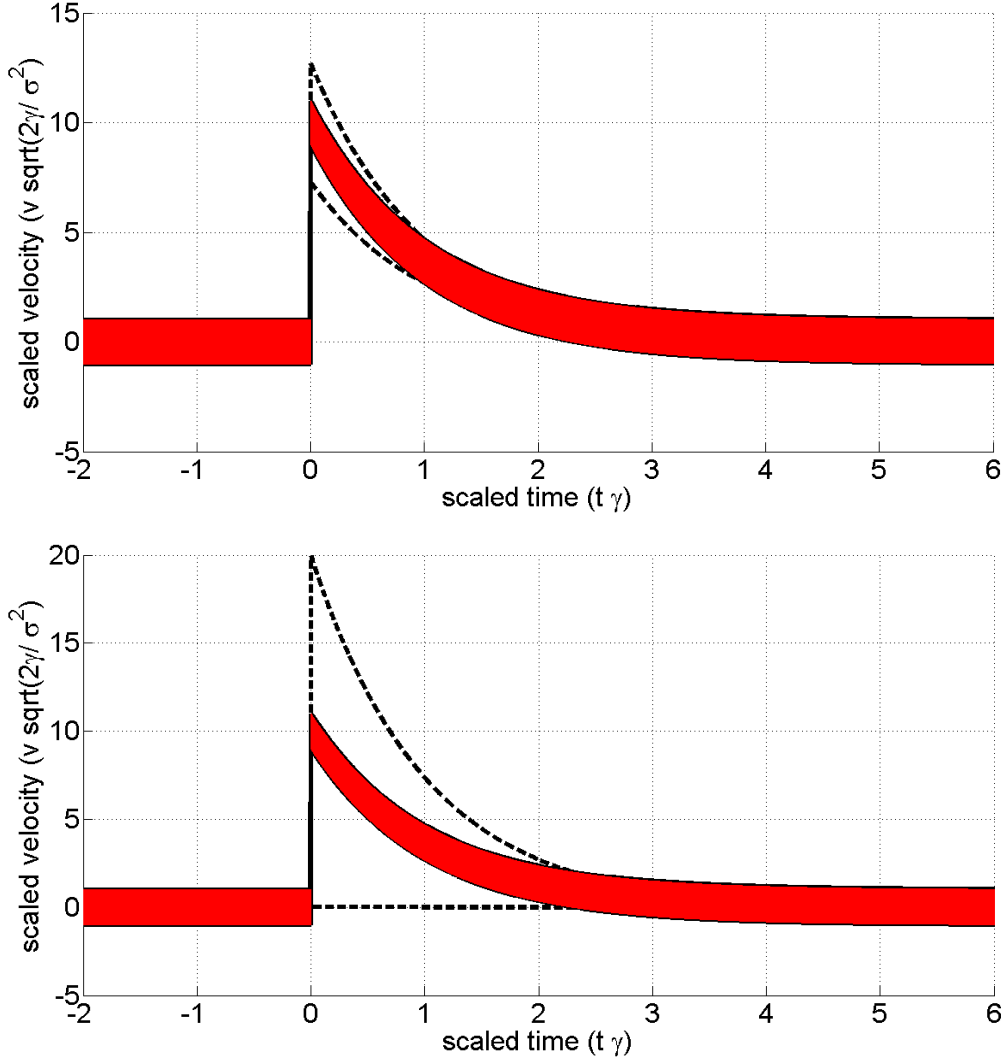

**Figure 5.** Reconstruction of kicking magnitude $v_0$ for the process shown in Figure 4 from one measurement at a scaled time $t\gamma = 1$ (top) and $t_{cut}\gamma = \ln(10) \approx 2.3$ (bottom). At this time $t_{cut}$ the mean value gets less than the standard deviation of the distribution $p_r$ of the reconstructed kicking magnitude and all the information about the kicking magnitude or even that a kick has occurred gets lost (see text).

The resolution $\delta_v$ according the Chernoff-Stein Lemma is given in (26) using $\frac{1}{2}m\bar{v}(t)^2$ as the mean value equation for the kinetic energy. In the remaining part of this subsection the resolution $\delta_v$ is compared to the standard deviation of the distribution for the reconstructed magnitude of the kick $p_r(v_r)$ in (12) (Figure 3). Using $f(v) = v \exp(-\gamma t)$ and $p_t$ from (21) we get from (12):

$$p_r(v_r) = \frac{1}{Z} \exp(-\gamma t) \exp\left(-\frac{\gamma}{\sigma^2}(v-v_0)^2 \exp(-2\gamma t)\right), \tag{27}$$

which is a Gaussian distribution for $v_r$ with mean $v_0$ and variance $Var(v_r) = \frac{k_B T}{m} e^{2\gamma t}$ (from (24)). The mean magnitude of the kick reconstructed by time reversal $v_r$ is $\bar{v}(t)\exp(\gamma t)$. If $n$ measurements at time $t$ of independent realizations of the process are given then $\bar{v}(t)$ is estimated by the arithmetic means of the measurement data, with $Var(v)/n$ as the variance. The distribution $p_r$ is Gaussian with a variance of

$$Var(v_r) = \frac{k_B T}{mn} e^{2\gamma t}. \tag{28}$$

Comparing with (26) one gets for the square of the resolution

$$\delta_v{}^2 = 2 Var(v_r) \ln\left(\frac{1}{\varepsilon}\right). \tag{29}$$

If the type II error $\varepsilon$ is set to $1/\sqrt{e}$, then $\delta_v{}^2 = Var(v_r)$. $\varepsilon$ is the error level in the Chernoff-Stein Lemma of estimating a kick magnitude $v_0 + \delta_v$ if the real kick was $v_0$. The resolution $\delta_v$ is proportional to the standard deviation of the reconstructed kick distribution. The factor depends only on the chosen error level $\varepsilon$.

As $\delta_v{}^2$ in (26) and $Var(v_r)$ in (28) scale with $1/n$, in (29) the number of measurements $n$ cancels out and (29) is valid not only for $n$ large but it is exact also for a small number $n$. For one realization of the process ($n=1$) the standard deviation, which is the square root of $Var(v_r)$, is shown in Figure 5 (top) for reconstructing the kick magnitude from a measurement at a scaled time $\gamma t = 1$. At a time of

$$\gamma t_{cut} := \ln\left(\frac{v_0}{\sqrt{Var(v)/n}}\right) \xRightarrow{t=t_{cut}} Var(v_r) = v_0{}^2 \tag{30}$$

the standard deviation of $v_r$ is $v_0$ (shown in Figure 5 (bottom) at $\gamma t_{cut} = \ln(10) \approx 2.3$). If we try to reconstruct the kick magnitude from measurement data acquired at a time $t > t_{cut}$ we cannot even recognize that a kick has occurred at $t = 0$, because the signal amplitude $\bar{v}(t)$ gets less than the fluctuations $Var(v)/n$. Therefore this criterion is used for the truncated singular value decomposition (SVD) method as the truncation criterion for the regularization of the inverse problem [2].

### 3.2. Brownian particle in a harmonic potential

In this subsection the Brownian particle in addition to stochastic damping proportional to the velocity $v$ is driven by a harmonic force proportional to $x$. Therefore the set of reduced variables which captures the information on the work consists of the two variables $x$ and $v$ and in addition to (17) also the second variable $x$ has to be described in the Langevin-equation:

$$\begin{aligned}\frac{dx(t)}{dt} &= v(t) \\ \frac{dv(t)}{dt} &= -\gamma v(t) - \omega_0{}^2 x(t) + \sigma\eta(t).\end{aligned} \tag{31}$$

The time-dependent distribution density, the mean values, and the (co)variances for $x$ and $v$ were presented already in 1943 by Chandrasekhar [47] for fixed initial values $x(0)$ and $v(0)$ at $t = 0$. Again we have changed the initial condition to the equilibrium state like for the free Brownian particle in subsection 3.1, having zero mean values with the canonical distribution density $p_{eq}(x,v) = \exp(-\beta H(x,v))/Z$ with $H(x,v) = \frac{1}{2}m\omega_0{}^2 x^2 + \frac{1}{2}mv^2$ is the sum of the potential and kinetic energy. At $t = 0$ a sudden shift in $x$ by $x_0$ and a kick in $v$ by $v_0$ results in the distribution $p_{kick}(x,v) = p_{eq}(x - x_0, v - v_0)$. As we have shown e.g. in [1] the distribution remains Gaussian for $t > 0$:

$$p_t(x,v) = \frac{1}{Z}\exp(-\beta H(x - \bar{x}(t), v - \bar{v}(t))) \tag{32}$$

where the mean values $\bar{x}(t)$ and $\bar{v}(t)$ are the solutions of the ordinary (non-stochastic) damped harmonic oscillator with initial values $x_0$ and $v_0$. Again for the equilibrium initial condition the variances stay constant in time:

$$Var(x) = \frac{k_B T}{m\omega_0^2}, \qquad Var(v) = \frac{k_B T}{m}, \tag{33}$$

and the covariance is zero. The same initial conditions have been used by Horowitz and Jarzynski [48], but for a Brownian particle in a dragged harmonic trap instead of a kick at $t = 0$. Gomez-Marin et al. [49] have applied for these initial conditions an instantaneous quench to the stiffness of the harmonic potential at $t = 0$ from $\omega_0$ to $\omega_1$, which gives time-dependent variances.

The damped harmonic oscillator can be reduced to an O-U process (subsection 3.1) in the overdamped limit, in which the velocity effectively equilibrates instantaneously, and the set of reduced variables, which captures the information on the work, is only the variable $x$. By setting in (31) the time derivative of the velocity to zero one gets

$$\gamma \frac{dx(t)}{dt} = -\omega_0^2 x + \sigma \eta(t), \tag{34}$$

which gives an exponential decay for the mean value $\bar{x}(t)$ with the time constant $\omega_0^2/\gamma$ and a variance $Var(x) = \frac{\sigma^2}{2\gamma\omega_0^2} = \frac{k_B T}{m\omega_0^2}$. This overdamped limit case has been often used in the past to present exact results on dissipation and fluctuation because mathematics is easier with one variable. For $x$ and $v$ as variables solutions are more complicated, but often do not give more insight in physical behavior [48]. For the presented cases we try to get successively to more complex systems with more variables. Therefore the full equation describing two variables including oscillations between potential and kinetic energy is important.

For the time-reversed process as sketched in Figure 3 we could use the known distribution $p_t(x, v)$ from (32) as the initial distribution and the time reversed equations of the damped harmonic oscillator to calculate distribution $p_r$ from (12). This distribution is Gaussian as $p_t(x, v)$ is Gaussian and the equations are linear. The standard deviation of $p_r$ could be used as a measure for resolution, as done in section 3.1 for the Ornstein-Uhlenbeck process. To avoid this rather lengthy calculation (compare Appendix of [48]) we can get the resolution easier from (15) using

$$\langle W_\delta(t) \rangle = \frac{1}{2} m\omega_0^2 \delta_x(t)^2 + \frac{1}{2} m \delta_v(t)^2 \approx \langle W_\delta \rangle \exp(-\gamma t). \tag{35}$$

with $\delta_x(t)$ and $\delta_v(t)$ are the solutions of the ordinary (non-stochastic) damped harmonic oscillator with initial values $\delta_x$ and $\delta_v$. Compared to the Ornstein-Uhlenbeck process the mean energy has not only an exponential decay in time with time constant $\gamma$ but for $\gamma < 2\omega_0$ also oscillations around that decay (Figure 6) [1], which explains the approximation in (35). The energy oscillates between potential and kinetic energy. Dissipation is the derivative of the mean total dissipated energy $\langle W_\delta(t) \rangle$ in Figure 6 and vanishes if the velocity and the kinetic energy gets zero.

If the type II error $\varepsilon$ is set to $1/\sqrt{e}$ (same as in section 3.1), the square of the resolutions in $x$ and $v$ are:

$$\delta_x^2 \approx \frac{k_B T}{m n \omega_0^2} e^{\gamma t}, \delta_v^2 \approx \frac{k_B T}{m n} e^{\gamma t}. \tag{36}$$

Up to now the two cases in section 3.1 and 3.2 had Gaussian distribution because the equations were linear. In the next section instead of a harmonic potential the Brownian particle is on the top of a pendulum and attracted in one direction by a force, e.g. gravity.

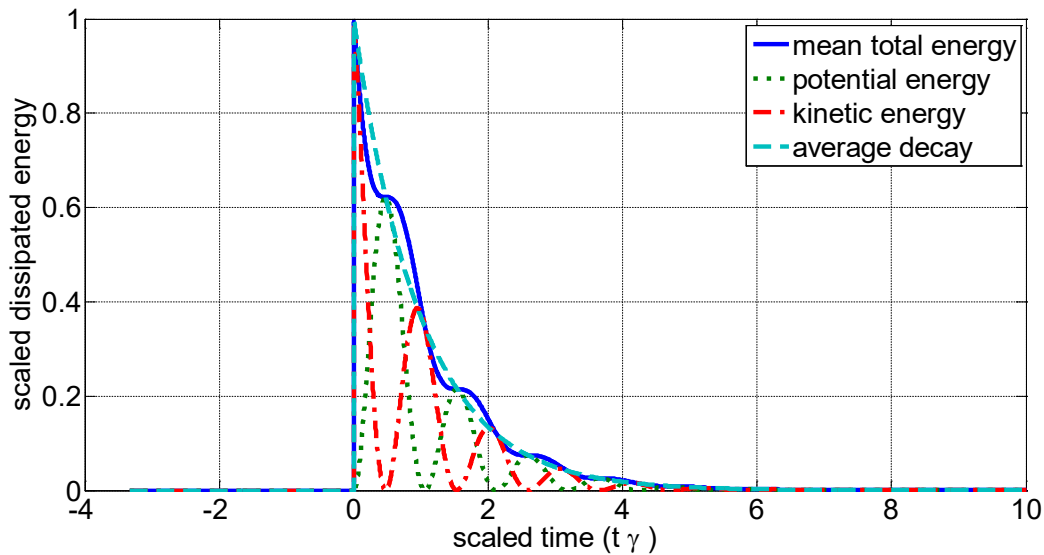

**Figure 6.** The mean total dissipated energy (solid line) of the damped harmonic oscillator is the sum of the potential energy (dotted) and the kinetic energy (dashdot) and shows in the average an exponential decay in time with time constant $\gamma$ (dashed line). For this figure the damping was chosen $\gamma = \frac{1}{3}\omega_0$.

### 3.3. Brownian pendulum in the overdamped limit

In this subsection the Brownian particle in addition to stochastic damping proportional to the velocity $v$ is driven by a force proportional to $\sin(x/l)$. $x$ can be thought as the path on the arc of a pendulum with length $l$. In the limit of a small amplitude of $x$ the Langevin-equation (31) with $\omega_0^2 = g/l$ is still valid, but for a higher amplitude the Langevin-equation is changed to:

$$\frac{dx(t)}{dt} = v(t)$$
$$\frac{dv(t)}{dt} = -\gamma v(t) - g\, sin\left(\frac{x}{l}\right) + \sigma\eta(t). \qquad (37)$$

In the overdamped limit, in which the velocity effectively equilibrates instantaneously, we get by setting the time derivative of the velocity to zero:

$$\gamma \frac{dx(t)}{dt} = -g\, sin\left(\frac{x}{l}\right) + \sigma\eta(t). \qquad (38)$$

The Hamiltonian is $H(x) = mgl\left(1 - \cos\left(\frac{x}{l}\right)\right)$. For the following numerical results we use $m = 1$, $\gamma = 1$, $\sigma = \sqrt{2}$ (which gives a variance of one for the linear case $x \ll l$), and $g = l$ and we get the Langevin-equation:

$$\frac{dx(t)}{dt} = -l\, sin\left(\frac{x}{l}\right) + \sqrt{2}\eta(t). \qquad (39)$$

For a higher parameter $l$, the equation gets more linear. If $l$ gets smaller, the non-linearity increases. This Langevin-equation (39) is equivalent to a description based on the Fokker-Planck equation for the time-dependent distribution density $p_t(x)$ [45]:

$$\frac{\partial p_t(x)}{\partial t} = \frac{\partial}{\partial x}\left(l\,sin\left(\frac{x}{l}\right) p_t(x)\right) + \frac{\partial^2 p_t(x)}{\partial x^2}. \tag{40}$$

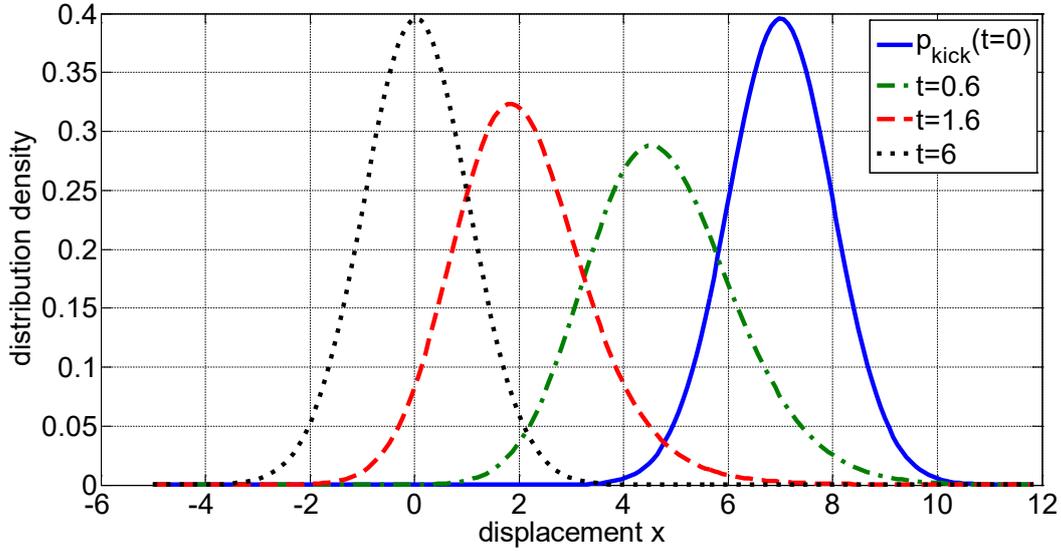

Figure 7. Distribution density $p_t$ for t=0 ($p_{kick}$), t=0.6, t=1.6, and t=6 (nearly already the equilibrium distribution $p_{eq}$) for $\beta = 1$ and $l = 4$ and a kick magnitude of $x_0 = 7$, numerically calculated from (40).

Again we start with an equilibrium $p_{eq}(x) = \exp(-\beta H(x))/Z$, which is no exact Gaussian distribution any more. Figure 7 shows the numerically calculated distribution density from (40) for $\beta = 1$ and $l = 4$: $p_{kick}$ for a kick magnitude of $x_0 = 7$, and $p_t$ for *t=0.6, t=1.6,* and *t=6* (nearly already the equilibrium distribution $p_{eq}$).

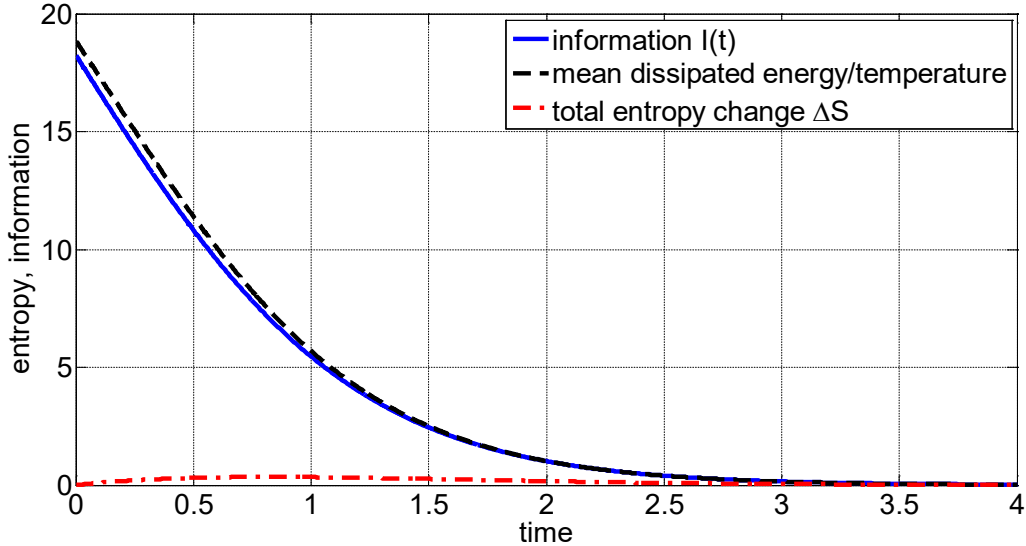

Figure 8. Information loss $I(t)$ and total entropy change $\Delta S$ after the kick for $\beta = 1$ and $l = 4$ and a kick magnitude of $x_0 = 7$. Even for rather "broad and overlapping" distributions the information can be approximated in (10) by the mean work, which has not been dissipated at time $t$, divided by the temperature: $H(x(t))/T$.

For the non-linear equation the shape of the distribution changes and the variance does not stay constant as for the linear equation. Therefore also the total entropy $\Delta S$ changes and is not zero all the time, but as shown in Figure 8, $\Delta S$ can be neglected compared to the information $I(t) = k_B D(p_t || p_{eq})$ and the first approximation in (10) is good despite the rather "broad and overlapping distributions" shown in Figure 7. In Figure 8 also the mean dissipated energy calculated as the Hamiltonian of the mean value $H(x(t))/T$ is shown and shows little deviation to $I(t)$. Therefore the second approximation in (10) is also adequate.

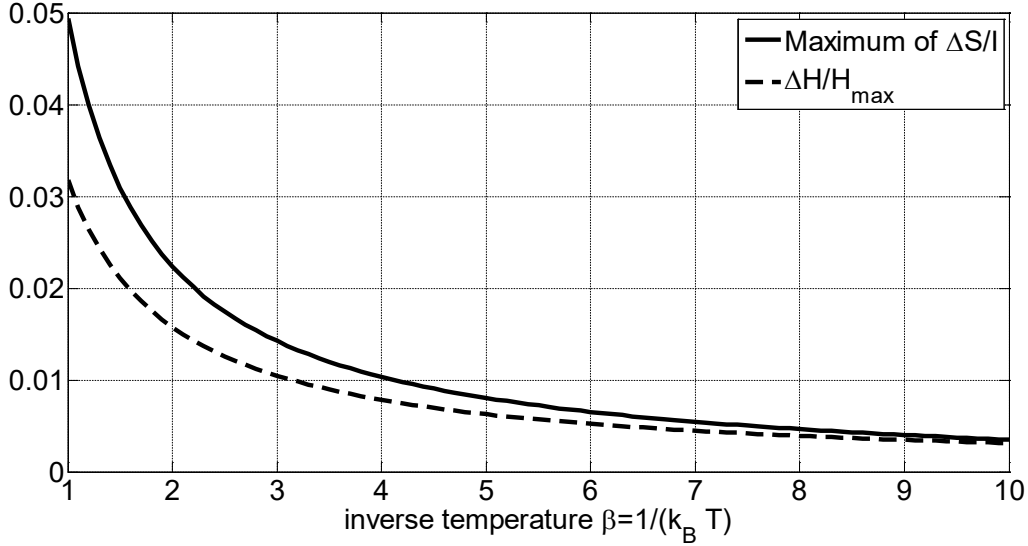

Figure 9. $\Delta S/I$ at the time when $\Delta S$ has its maximum as a function of the inverse temperature $\beta$. For "smaller" distributions, which can be realized by a lower temperature (higher $\beta$) the first approximation in (10) gets even better and is practically exact for room temperature. Also the second approximation in (10) is good which can be seen by $\Delta H / H_{max} \ll 1$ (see text).

For "smaller" distributions, which can be realized by a lower temperature (higher $\beta$), the approximations get even better. Figure 9 shows the maximum of $\Delta S(t)$ divided by $I(t)$ and $\Delta H/H_{max}$ with $\Delta H = H_{max} - (\langle H \rangle_{kick} - \langle H \rangle_{eq})$ and $H_{max} = (H(x_0) - H(x = 0))$ as a function of the inverse temperature $\beta$. For the Hamiltonian of the Brownian pendulum $\Delta H/H_{max} = \left(1 - \langle \cos\left(\frac{x}{l}\right)\rangle_{eq}\right)$, which can be derived by using the angle sum formula for the cosine.

For the *inverse problem* of estimating the distribution $p_r$ of the kicked magnitude $x_r$ from the distribution $p_t$ at a certain time $t > 0$ after the kick (compare (12)), the *time reversal* of the Langevin-equation (39) without the noise term or of the equivalent Focker-Planck equation (40) without the diffusion term can be used:

$$\frac{\partial p_t(x)}{\partial t} = -\frac{\partial}{\partial x}\left(l \sin\left(\frac{x}{l}\right) p_t(x)\right). \tag{41}$$

Starting with the initial distribution $p_t$ and numerically solving (41) one gets the distribution $p_r$ after the time $t$. This has been performed for $\beta = 50$, $l = 4$ and $t = 2$ in Figure 10: $p_{kick}$ is shown for a kick magnitude of $x_0 = 5.5$ ($\delta = 0$) and $x_0 + \delta = 6$ ($\delta = 0.5$). Also the distribution $p_t$ after a time $t = 2$ and the reconstructed distribution $p_r$ are shown for both kick magnitudes ($\delta = 0$ and $\delta = 0.5$, compare Figure 2 and Figure 3). The standard deviation of the reconstructed distribution $p_r$ can be taken as a measure of the spatial resolution: $std(p_r(\delta = 0)) \approx 0.6$. For later times than $t = 2$ or for more "peaked" distributions (e.g. by a higher $\beta$) the direct numerical calculation from (41) of the reconstructed distribution gets unstable as it is an ill-posed problem, but the resolution can still be estimated by using the Chernoff-Stein Lemma as described in (14) and (15).

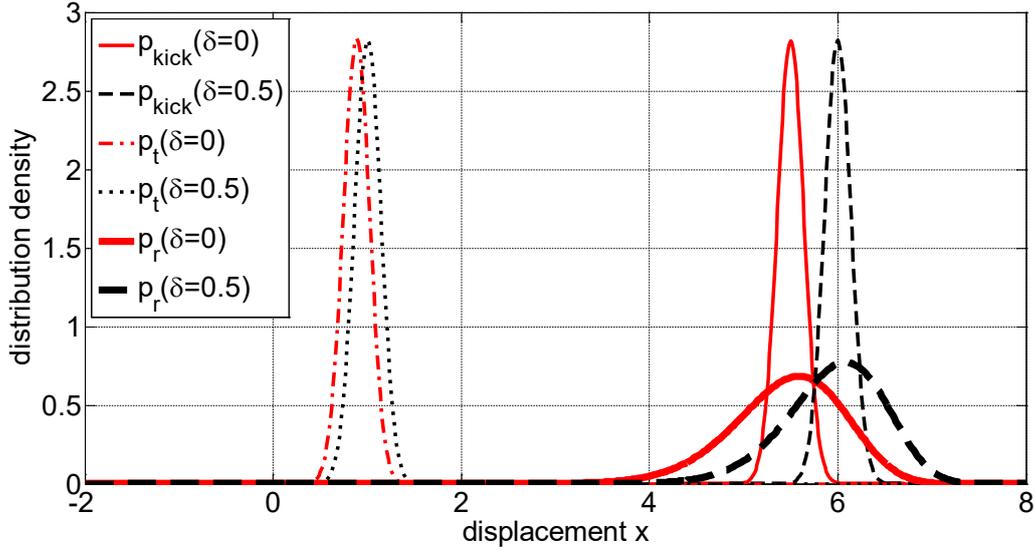

Figure 10. Distribution densities $p_{kick}$, $p_t$ after a time $t = 2$ and the reconstructed distribution $p_r$ are shown for both kick magnitudes $x_0 = 5.5$ ($\delta = 0$) and $x_0 + \delta = 6$ ($\delta = 0.5$).

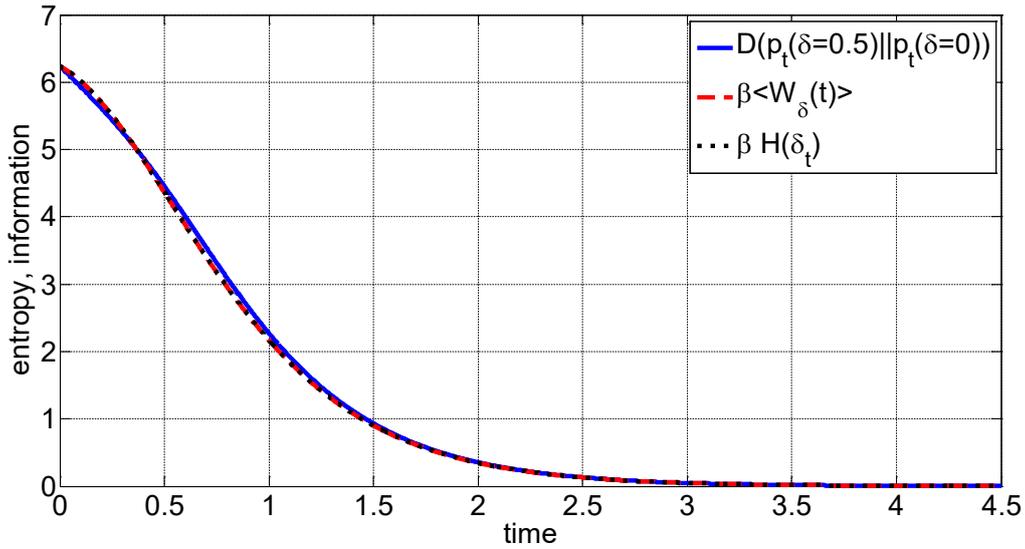

Figure 11. The distance $D(p_t(\delta = 0.5)||p_t(\delta = 0))$ for the initially ($t=0$) kicked distributions with $x_0 = 5.5$ ($\delta = 0$) and $x_0 + \delta = 6$ ($\delta = 0.5$) decreases with time and is similar to the dissipated energy $\beta\langle W_\delta(t)\rangle$ and $\beta H(\delta_t)$.

In Figure 8 according to equation (10) the decreasing "distance" $D(p_t||p_{eq})$ between the distribution $p_t$ and the equilibrium distribution $p_{eq}$ is shown. Similarly in Figure 11 according to (14) the distance between the distribution $p_t$ for a kick magnitude $x_0 + \delta = 6$ ($\delta = 0.5$) and $x_0 = 5.5$ is shown. To check the approximations in (14) also $\beta\langle W_\delta(t)\rangle$ and $\beta H(\delta_t)$ are shown. According to (13) the first approximation is exact for $t=0$ with $D(p_{kick}(\delta = 0.5)||p_{kick}(\delta = 0)) = \langle W_\delta \rangle \approx 6.238$. After a long time this distance becomes zero as equilibrium is reached. For an intermediate time the approximations in (14) get better for smaller distributions, which is the case for macroscopic systems (compare Figure 9). After the time $t = 2$, where $p_t(\delta = 0)$ and $p_t(\delta = 0.5)$ are overlapping as shown in Figure 10, the distance of the mean values is reduced to $\delta_t \approx 0.1181$ and $D(p_t(\delta = 0.5)||p_t(\delta = 0)) \approx 0.356$, as shown in Figure 11. For comparison $\beta\langle W_\delta(t = 2)\rangle \approx 0.3486$ and $\beta H(\delta_t) \approx 0.3488$, which is a good approximation for distance between the two distributions measured by $D(p_t(\delta = 0.5)||p_t(\delta = 0))$.

The distance of the mean values $\delta_t$ and therefore $\beta H(\delta_t)$ can be calculated from the mean value equations, which is for macroscopically "thin" distributions (Figure 1) the same equation as the Langevin-equation (39), but without the noise term. As $\delta_t$ is small compared to the amplitude of the Brownian pendulum one gets in first order approximation:

$$\frac{d\delta_t}{dt} = -\delta_t \cos\left(\frac{x(t)}{l}\right), \quad (42)$$

where $x(t)$ is the solution of the macroscopic mean value equation. This is an essential result: the macroscopic mean value equation determines $\beta H(\delta_t)$, which is a good approximation for $\langle W_\delta(t)\rangle$ and the KLD $D(p_t(\delta)||p_t(\delta = 0))$ as a property of the statistical fluctuations (14). Using the Chernoff-Stein Lemma (15) for a fixed resolution $\delta$ the error $\varepsilon$ of taking the wrong distribution can be estimated as a function of time – or if inverted – for a fixed error $\varepsilon$ the resolution $\delta_r(t)$ can be estimated, which is shown in Figure 12 for $l = 4, 5,$ or 8 for a kick magnitude of $x_0 = 5.75$ (= $x_0 + 0.5\,\delta$ from Figure 10). Error $\varepsilon$ is set to $1/\sqrt{e}$ to get for the resolution the same as the standard deviation of the reconstructed distribution $p_r$ (compare $\delta_r(t = 2) \approx 0.6$ for $l = 4$ in Figure 10). In Figure 12 the used solution of the macroscopic mean value equation $x(t)$ is also shown as well as $x_0 \exp(-t)$ as the exponential decay for the linear equation in (17) for the limit of large $l$. The time $t_{cut}$ has been defined for the linear equation in (30) as the time when the resolution is equal to the kick magnitude $x_0$, which gives $\ln(x_0) \approx 3.7$ (dashed line in Figure 12). For increasing non-linearity (decreasing $l$) the cut-off time $t_{cut}$ increases. From the point of the Chernoff-Stein Lemma the cut-off time $t_{cut}$ is the time when the distribution $p_t$ cannot be distinguished from the equilibrium distribution $p_{eq}$. According to (10) $D(p_{t_{cut}}||p_{eq})$ can be approximated by $\beta H(x(t_{cut}))$, which - using the Chernoff-Stein Lemma (15) - can be approximated by $\frac{1}{n}\ln\left(\frac{1}{\varepsilon}\right)$.

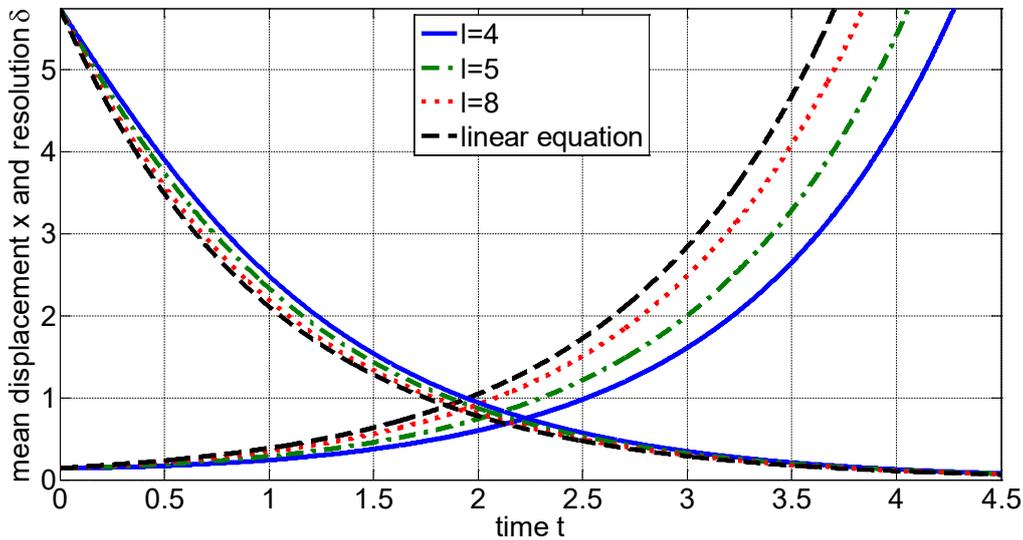

Figure 12. Mean value decay of the displacement $x(t)$ and increasing resolution $\delta_r(t)$ for $l = 4, 5,$ or 8 and a kick magnitude of $x_0 = 5.75$. The dashed line shows the exponential decay of the displacement and the increasing resolution for the linear equation in (17) for a large $l$.

## 4. Summary, conclusions and outlook on future work

In macroscopic systems fluctuations, which are deviations from the average behavior, get small and can be neglected in the thermodynamic limit. But for inverse problems these fluctuations are highly "amplified", which is shown in section 2 for a system kicked out of the equilibrium, followed by a dissipative process back to equilibrium (Figure 1). The inverse problem of estimating the kick-magnitude from an intermediate state a certain time after the kick is ill-posed. Just after the kick its magnitude can be estimated very well. A long time after the kick the state is nearly dissipated back to equilibrium and nearly all the information about the kick magnitude is lost. For macroscopic systems it could be shown in section 2 that the information loss is in a good approximation just the mean dissipated energy divided by the temperature, which is the mean entropy production. In imaging, spatial resolution and information content are strongly correlated, and therefore a loss of information results in a loss of resolution, which is quantified in section 2.5.

In section 3 the mean dissipated energy and the resulting loss of resolution is calculated for a kicked Brownian particle, either free or driven by a linear or non-linear force. For these study cases also the time-dependent probability distribution could be explicitly calculated and the deduced information loss and the resolution are in excellent agreement with the results from mean value calculations, even when the distributions are broad and still far away from the thermodynamic limit. This confirms the approximations made for macroscopic systems as described in section 2. A cut-off time could be given, for which the state after that time cannot be distinguished from the equilibrium distribution according to the Chernoff-Stein Lemma, when the Kullback-Leibler divergence gets too small.

To conclude, for macroscopic systems it is not necessary to describe the full stochastic process to get the influence of the fluctuations. The relevant information loss can be calculated from the averaged behavior (mean value), which describes the usually known macroscopic evolution of the system in time, as the mean dissipated work divided by the temperature. This remarkable feature might be the reason that regularization methods for ill-posed inverse problems work so well, although they use only the mean value equations and not the detailed stochastic process to describe the time evolution. The choice of an adequate regularization parameter, e.g. the cut-off value for the truncated singular value decomposition (SVD) method, is equivalent to the choice of the error level $\varepsilon$ in the Chernoff-Stein Lemma [2].

A prominent example of an ill-posed inverse problem is non-destructive imaging, where the information about the spatial pattern of a sample's interior has to be transferred to the sample surface by certain waves, e.g., ultrasound or thermal waves. Imaging is done by reconstruction of the interior structure from the signals measured on the sample surface, e.g., by back-projection or time-reversal for a photo-acoustically induced ultrasound wave [50, 51]. There are several effects which limit the spatial resolution for photoacoustic imaging. Beside insufficient instrumentation and data processing one limitation comes from attenuation of the acoustic wave. This information loss during the acoustic wave propagation cannot be compensated by any signal processing algorithm. We have tried to compensate this degradation of spatial resolution by using regularization methods, and it turned out that thermal fluctuations limit the spatial resolution [1]. In the present work it is shown that the loss of information, which is equal to the entropy production as the dissipated energy divided by the temperature, is a principle thermodynamic limit, which cannot be compensated. This is also true for heat diffusion as the mean value equation used for thermographic imaging. In the past we have modeled heat diffusion by a Gauss-Markov process in Fourier space and found a principle limit for the spatial resolution [2]. Using the information loss and entropy production for a kicked process derived in section 2 it is shown that the resolution limit depends just on the macroscopic mean-value equations and is independent of the actual stochastic process, as long as the macroscopic equations describe the mean work and therefore also the mean dissipated work.

In future work actual thermodynamic resolution limits for photoacoustic or thermographic imaging should be given by using the entropy production in an attenuated acoustic wave or for thermal diffusion, respectively [52]. The waves can be represented as a superposition of wave trains having a certain wavenumber or frequency in Fourier k-space or ω-space. The mean entropy production gives a criterion for a cut-off wavenumber or a cut-off frequency, where all the information about the Fourier-component is lost because it cannot be distinguished from equilibrium according the Chernoff-Stein Lemma. In addition to a short kick also other excitation patterns can be considered to evaluate the resolution limits. G. Busse used a phase angle measurement with a sinusoidal excitation in lock-in thermography to get a better resolution in depth [53]. Sreekumar and Mandelis proposed a chirped excitation pattern like in radar technology to get a better spatial resolution at a certain depth [54]. Using the mean entropy production it should be possible to give also thermodynamic resolution limits for those excitation patterns and compare them to resolution limits for single short pulse excitation.

**Acknowledgments**

It is a pleasure to thank Juan MR Parrondo for valuable discussions. This work has been supported by the Austrian Science Fund (FWF), project number S10503-N20, by the Christian Doppler Research Association, the Federal Ministry of Economy, Family and Youth, the European Regional Development Fund (EFRE) in the framework of the EU-program Regio 13, and the federal state of Upper Austria.